\documentstyle[epsfig]{aipproc}

\begin{document}

\title{Starbursts: Lessons for the Origin and Evolution of Galaxies
and the Inter-Galactic Medium}

\author{Timothy M. Heckman}

\address{$^*$ Department of Physics \& Astronomy\\
Johns Hopkins University\\ Baltimore, Maryland 21218\\
$^{\dagger}$ Adjunct Astronomer, Space Telescope Science Institute}

\maketitle

\begin{abstract}
Starbursts are episodes of intense star-formation that occur in the
central regions of galaxies, and dominate the integrated emission
from the galaxy. They are a significant component of the present-
day universe, being the site of $\sim$ 25\% of the high-mass star-
formation. They offer unique `laboratories' for testing our ideas
about star-formation, the evolution of high-mass stars, and the
physics of the interstellar medium. They serve as local analogs of
the processes that were important in the origin and early evolution
of galaxies and in the heating and chemical enrichment of the
inter-galactic medium. In this contribution I review starbursts
from this broad cosmogonical perspective, stressing several key
lessons we have learned from starbursts: 1) Violent, transient
events play a significant role in the origin and evolution of
galaxies. 2) Galaxies do not evolve as `Island Universes':
starbursts are triggered by galaxy interactions and produce
outflows of hot chemically-enriched gas that `pollute' the inter-
galactic medium. 3) Dust dramatically affects of view of high-mass
star-formation in starbursts and (probably) in high-redshift
galaxies. Throughout this review I emphasize the importance of
space-based observations in understanding starbursts.
\end{abstract}

\section*{Introduction}

\subsection*{What is a Starburst?}

Starbursts are brief episodes of intense star-formation that occur
in the central-most 10$^{2}$ to 10$^{3}$ pc-scale regions of
galaxies, and dominate the overall luminosity of the galaxy. Their
star-formation rates are so high that the existing gas supply can
sustain the starburst for only a small fraction of the age of the
universe (in agreement with detailed models of the observed
properties of starbursts, which imply typical burst ages of-order
10$^{8}$ years or less). Comprehensive reviews of the starburst
phenomenon may be found in the volumes edited by Leitherer et al
(1991) and Franco (1997).

Starbursts are powered by high-mass stars, which I will define to
have masses greater than 8 M$_{\odot}$. These stars are very
luminous (L $>$ few thousand L$_{\odot}$), and so (when present in
significant numbers) they dominate the energy output of a
population of stars. They are also short-lived (a few million to
a few-tens-of-million years) and therefore trace relatively recent
star-formation. High-mass stars dominate the heating of the
interstellar medium, the gas out of which new stars form. They are
also the `thermonuclear furnaces' that forge most of the elements
heavier than He, and then disperse these heavy elements (hereafter
`metals') back into the interstellar medium when the dying high-
mass stars explode as supernovae.

In the local universe, high-mass stars are found in both normal
galaxies like our own Milky Way and in starburst galaxies. In
normal galaxies they are distributed in the spiral arms throughout
the $\sim$ 30-kpc-scale disk of the galaxy. In such galaxy disks,
the existing supply of gas can sustain the current rate of star-
formation for many Gigayears, consistent with a relatively slow
evolution of the star-formation rate over much of the history of
the universe. As already stated above, the high-mass stars formed
in starburst galaxies are concentrated into a small region in the
galaxy center ($\sim$ 100 times smaller than the galaxy-as-a-whole)
and are the consequence of a transient event with a duration $<$
10$^{8}$ years (i.e. $<$ 1\% the present age of the universe).

High-mass stars are typically very hot (T $>$ 25,000 K), and thus
the luminous output of a starburst should have its peak in the
vacuum-ultraviolet range ($\lambda$ $\sim$ 912 to 3000 \AA). On the
other hand, starbursts are rich in interstellar gas and dust. The
effective absorption cross-section of these dust grains is a strong
inverse function of wavelength (cf. Calzetti 1997). The dust grains
therefore absorb much of the space-ultraviolet starlight and are
thereby heated to temperatures of-order 10 to 100 K. The grains
then cool by emitting far-infrared radiation. Thus, samples of low-
redshift starburst galaxies have come mostly from surveys that
select galaxies with unusually bright ultraviolet or far-infrared
emission (cf. Huchra 1977; Soifer et al 1989).

\subsection*{Why are Starbursts Important?}

First of all, starbursts are a very significant component of our
present-day universe, and as-such deserve to be understood in their
own right. The most complete and well-characterized sample of
starbursts is that drawn from the far-infrared IRAS survey (e.g.
Soifer et al 1989). The rather strong inverse correlation between
gas-depletion timescales and luminosity means that we may consider
that starbursts dominate the set of galaxies lying above the `knee'
in the far-infrared galaxy luminosity function (L$_{IR} >$ few $\times
10^{10}$ L$_{\odot}$). 
By this reckoning,
starbursts provide about 10\% of the bolometric emissivity of the
local Universe. Thus, starbursts are an energetically-significant
phenomenon.

It is also instructive to estimate the rate of high-mass star-
formation in starbursts compared to that in the disks of normal
galaxies in the local universe. High-mass stars are so hot and
luminous that they are the dominant source of photons energetic
enough to photoionize Hydrogen in the interstellar medium. As the
ionized hydrogen recombines and the captured electron cascades to
the ground-state, it emits photons. Thus, the luminosity of these
Hydrogen emission-lines is a measure of the total number of high-
mass stars in a galaxy. Since we know the lifetimes of these stars,
this yields an average rate at which such stars must have formed
in the recent past (cf. Kennicutt 1983; Leitherer \& Heckman 1995).
Of course the observed fluxes of the Hydrogen emission-lines must
be corrected for the effects of extinction by dust. In normal
galaxies (where the extinction due to dust is not severe), the optical
Balmer emission-lines can be used. In dusty starbursts, infrared or mm-
wave lines (which are much less affected by dust-extinction) must
be used.

Applying this technique to the Kraan-Korteweg \& Tammann (1979)
volume-limited catalog of galaxies in the nearby universe (D $<$
10 Mpc), I find that the four most luminous circum-nuclear
starbursts in the local universe (M 82, NGC 253, M 83, and NGC
4945) together comprise about 25\% of the total high-mass star-
formation in this volume, and the six most-actively-star-forming
disks of normal galaxies contribute a comparable amount. In
specific terms, the rate of high-mass star-formation in the few-
hundred-parsec-scale starburst in M 82 actually exceeds the rate
in the entire disk of the spiral galaxy M 101 (which has a surface
area approximately four orders-of-magnitude larger than the M 82
starburst)! Thus, both in terms of total energy production and rate
of high-mass star-formation, starbursts are indeed highly
significant components of the present universe.

While starbursts are therefore fascinating and significant objects,
they are even more important when placed in the broader context of
contemporary stellar and extragalactic astrophysics. They are ideal
laboratories in which to study: 1) The formation and evolution of
massive stars 2) The physics of the interstellar medium under
extreme conditions 3) The processes involved in the formation and
early evolution of galaxies 4) The processes likely responsible for
chemically-enriching and heating the intergalactic medium.

The cosmological relevance of starbursts has been dramatically
underscored by one of the most spectacular discoveries in years:
the existence of a population of high-redshift (z $>$ 2) field
galaxies (cf. Steidel et al 1996; Lowenthal et al 1997; Pettini,
this volume). The  number density of these galaxies implies that
they almost certainly represent precursors of typical present-day
galaxies in an early actively-star-forming phase. This discovery
moves the study of the star-forming history of the universe into
the arena of direct observations (cf. Madau et al 1996; Madau, this
volume), and gives added impetus to the quest to understand local
starbursts.

It is these aspects of the starburst phenomenon - the way in which
nearby starburst galaxies can used to address mAJor issues in
cosmogony - that are the subject of the rest of this contribution.

\subsection*{What lessons have we learned from starbursts?}

Before discussing starbursts in more detail, it is useful to
briefly summarize the mAJor lessons in extragalactic astronomy and
cosmology that we have learned from the study of starbursts in the
local universe.\\[1mm]

{\bf i) Violent, transient events play a significant role in the
origin and evolution of galaxies.}

Current interest in starbursts has been stimulated by (and has
helped to create) a paradigm shift in which the importance of
cataclysmic events in the evolution of galaxies has become
increasingly recognized. Rather than simply evolving in a steady
`clockwork' fashion, galaxies may also evolve discontinuously
through bursts of star-formation. This is reminiscent of the
concept of `punctuated equilibrium' for the evolution of biological
systems.\\[1mm]

{\bf ii) Galaxies do not evolve as `Island Universes' or `Closed
Boxes.'}

There is instead two-way causal communication between galaxies and
their environment: strong starbursts are triggered by the
interaction or merger of two galaxies and the subsequent starbursts
produce outflows of ionizing radiation and metal-enriched matter
that travel into the galactic halos and possibly beyond. Further
pursuing the biological analogy, this might be thought of as the
`ecology' of galaxy evolution.\\[1mm]

{\bf iii) Dust in galaxies drastically affects our view of star-
formation in local starbursts and (most likely) in galaxies in the
early universe.}

As noted earlier, the intrinsic spectral energy distribution of
a young stellar population peaks in the vacuum-ultraviolet, just where the
opacity of the interstellar dust grains that permeate the 
region of star-formation is
maximal. This effect is especially relevant at high-redshifts,
since the visible-light observations that provide the bulk of our
information about these galaxies sample the vacuum-ultraviolet
portion of their rest-frame spectrum.

\section*{The `Ecology' of Starbursts}

\subsection*{The Causes of Starbursts}

Starburst galaxies require the presence of substantial amounts of
gas within the central-most few hundred parsecs of the galaxy. The
relatively low efficiency of stars for
energy production (E $\sim10^{-3}$ M$_{star}$c$^{2}$), coupled with the
severe energetic demands (ranging from $\sim$10$^{59}$ ergs for a
modest starburst like M 82 up to $\sim$ 10$^{61}$ ergs for an
`ultraluminous' starburst like Arp 220) mean that interstellar gas
masses of at least 10$^{8}$ to 10$^{10}$ M$_{\odot}$ are needed to
`fuel' a starburst (even assuming 100\% efficiency for the
conversion of gas into stars). Indeed, mm-wave interferometric maps
of the molecular gas in powerful starbursts provide direct
observational evidence for such material (cf. Sanders \& Mirabel
1997).

As Larson (1987), Elmegreen (1997), and others have argued, the
surface mass density of the cold interstellar matter in starburst
nuclei is so high (typically of-order 10$^{3}$ M$_{\odot}$ pc$^{-
2}$), that the growth time for gravitational instabilities that
would lead to star-formation in the starburst is extremely short
($\sim$ a million years). Moreover, the timescales for gas
depletion in a starburst via star-formation and/or supernova-driven
outflows (see section 3.2 below) are also short compared to the
minimum time it would take gravity to move material into the
starburst from the large-scale disk of the galaxy (10$^{7}$ to
10$^{8}$ years vs. 10$^{8}$ to 10$^{9}$ years respectively). Larson
therefore argues that since the gas that fuels the starburst must
be assembled at least as fast as it is consumed, and since the gas
has a mass comparable to the entire mass of the interstellar medium
in a normal galaxy, powerful starbursts can only occur when some
process allows a substantial fraction of the interstellar medium
of a galaxy to flow inward by at least an order-of-magnitude in
radius at velocities that are comparable to the orbital velocities
in the galaxy's disk.

By way of illustration, Larson emphasizes that the collapse of a
self-gravitating system implies a maximum infall rate of roughly
25 (v$_{infall}$/50 km s$^{-1}$)$^{3}$ M$_{\odot}$ per year. This
can be compared to typical estimated star-formation rates of 10 to
100 M$_{\odot}$ per year in starbursts and typical orbital
velocities of roughly 150 km s$^{-1}$ in the `host' galaxy of the
starburst (e.g. Lehnert \& Heckman 1996b). To summarize, the
fueling of a starburst requires a mechanism that can induce non-
circular motions that are both large in amplitude and involve a
substantial fraction of the interstellar medium of the galaxy. 

Fortunately, these theoretical or heurstic arguments are supported
by observational evidence. As samples of increasingly luminous
starbursts are examined, the strength of the evidence that they are
strongly interacting or merging systems also increases (cf. Sanders
\& Mirabel 1997). Such evidence is mostly
morphological: 1) The presence of two or more nuclei within a
single distorted envelope (consistent with a nearly-complete merger
of two galaxies). 2) The presence of a second galaxy with the
requisite proximity and relative brightness, together with the
bridges and long linear `tails' that are the hallmark of tidally-
interacting galactic disks (Toomre \& Toomre 1972).

The physical picture that has emerged from the theoretical
interpretation of inceasingly-sophisticated numerical simulations
(cf. Mihos \& Hernquist 1994a,b) is that during the close passage
of two galaxies, tidal stresses act to strongly perturb the orbits
of the stars and gas in the galaxy disk. The dissipation of kinetic
energy as gas collides with gas, allows the gas to become
sufficiently displaced from the stars that gravitational torques
act between the stars and gas to transfer significant amounts of
angular momentum from the gas to the stars. The gas can thereby
fall into the center of the galaxy, where it can fuel a starburst.
If the passage of the two galaxies is slow and inter-penetrating
enough, dynamical friction can transfer enough kinetic energy from
the stars to the dark-matter halo to allow the two galaxies to
merge into a single galaxy.

Such mergers or strong interactions should take a few times the
galaxy rotation period (e.g. $\sim10^{9}$ years), with the intense
starburst phase being significantly shorter (cf. Mihos \& Hernquist
1994a,b). These timescales are loosely consistent with independent
estimates of starburst lifetimes, and (in any case) are quite
consistent with the transient nature of starbursts.

\subsection*{The Byproducts of Starbursts}

Perhaps the most spectacular aftermath of a starburst is the
`life-changing' event associated with a galaxy merger. As first
suggested by Toomre (1977), it is now clear both observationally
and theoretically that the merger between two disk galaxies is at
least one avenue for producing an elliptical galaxy (cf. Schweizer
1992; Barnes 1995). The presence of `kinematically-decoupled' cores
in elliptical galaxies (central regions in which the stars have a
very different angular momentum axis from the rest of the galaxy)
may be a fossil record of an merger-induced starburst in the
distant past (e.g. Franx \& Illingworth 1988). Likewise, the so-
called `E+A' galaxies, whose spectra appear to be the sum of the
stars found in normal elliptical (E) galaxies plus a population
dominated by A stars (stars having a lifetime of 0.1 to 1
Gigayears), may be `post-starbursts' systems that are intermediate
in evolutionary state between merger-driven starbursts and bona-
fide elliptical galaxies (cf. Zabludof et al 1997).

Another very intruiging consequence of starbursts is the `super
star cluster' phenomemon. Meurer et al (1995) find that typically
20\% of all the vacuum-ultraviolet light (and hence $\sim$20\% of
the high-mass stars) in starbursts is produced by very luminous
stellar clusters (dubbed `super star clusters' by O'Connell et al
1994). These clusters are so luminous that it is very tempting to
describe them as the possible progenitors of globular clusters (cf.
Whitmore \& Schweizer 1995). While the relationship between the super star
clusters and classical globular clusters is a matter of on-going
debate (cf. van den Bergh 1995; Meurer 1995), work on the nearest
such objects demonstrates that they indeed have sizes, velocity
dispersions, and masses that are similar to those of typical
globular clusters (Ho \& Filippenko 1996a,b).

More generally, the census of high-mass star-formation in the local
universe described in section 1.2 above has an very interesting
implication. Adopting the `Copernican Principle' (that is, assuming
that our local piece of the present-day universe is representative
of the universe elsewhere and elsewhen), the fact that $\sim$25\%
of all high-mass stars are formed in starbursts would mean one of
the following:\\[1mm]

{\bf i) Operating over the history of the universe, starbursts make
about 25\% of {\it all} the stars in galaxies (not just high-mass
stars).} Given the central location of starbursts in galaxies and
the fact that the most powerful starbursts seem to be associated
with the building of elliptical galaxies via mergers, the only
plausible fossil record of these ancient starbursts would be the
central parts of disk galaxies (bulges and inner disks) and
elliptical galaxies. However, bulges and ellipticals are redder and
older than the average disk of a spiral galaxy, so it may be
difficult to make this idea hang-together quantitatively. That is,
starbursts may have indeed produced $\sim$25\% of the stars in
galaxies on-average integrated over the history of the universe,
but could not be doing-so today or the central parts of galaxies
would be too blue and too bright.\\[1mm]

{\bf ii) Starbursts (unlike normal galactic disks) make only high-
mass stars and therefore leave little long-term residue.} By this
I mean that high-mass stars are short-lived ($<$ few $\times10^{7}$
years) compared to low-mass stars (e.g. 10$^{10}$ years for the
sun). If starbursts form only high-mass stars, then shortly after
the burst is over, the starburst would fade away completely,
leaving only a residue of neutron stars and black holes (the
`Chesire Cat' model). Leitherer (1997) reviews the evidence that
starbursts may `manufacture' only high-mass stars (cf. Rieke et al
1993 vs. Satyapal et al 1997). In contrast, star-formation with a
normal initial mass-function extending down to well below 1
M$_{\odot}$ produces stars (and hence starlight and stellar mass)
that remain behind essentially forever ($>$ the age of the
universe). This is in fact how we believe the large-scale disks of
normal galaxies are constructed over the course of many Gyrs.

\section*{Starbursts and the Inter-Galactic Medium}

The Inter-Galactic Medium (`IGM') represents the mAJority of the
baryonic matter in clusters of galaxies (cf. Donahue, this volume).
Comparison of the total baryonic content of the universe (cf.
Turner et al 1996) to the baryons located inside the visible parts
of galaxies implies that perhaps as much as 80 to 90\% of the
baryons in the universe might reside in the IGM. The source for
the heating and re-ionization of the IGM at high-redshifts is a
puzzle, since the IGM is already ionized at the time the first
known QSOs appear at z $\sim$ 5. Ionization due to a pre-QSO phase
of galaxy formation has been suggested as possibility (cf. Madau
\& Shull 1996; Miralda-Escude \& Rees 1997; Giroux \& Shapiro
1996). As I will argue below, starbursts may play a vital role not
only in heating the IGM, but also in `polluting' it with the heavy
elements forged by high-mass stars.

\subsection*{The Photoionization of the IGM by Starbursts}

The high-mass stars in starbursts may be a copious source of
photons that are energetic enough to ionize Hydrogen and neutral
Helium in the IGM. Put most simply, for star-formation with a
normal initial-mass function that extends down to 0.1 M$_{\odot}$
(8 M$_{\odot}$), each baryon that is incorporated into stars yields
the emission of roughly 3000 (20000) photons that are capable of
ionizing Hydrogen. Thus, a little bit of star-formation could go
a long way in photoionizing the IGM (especially during very early
times before the earliest known QSOs appear on the scene).

However, direct observations of local starburst galaxies below the
Lyman limit at 912\AA\ using the Hopkins Ultraviolet Telescope show
that only a small fraction of the total ionizing radiation produced
by these starbursts (as measured by the luminosity of the Hydrogen
recombination-lines) escapes into the IGM (Leitherer et al 1995;
Hurwitz et al 1997). Giallongo et al (1997) have recently set an
upper limit of 20\% on the average fraction of ionizing radiation
due to high-mass stars escaping from all galaxies out to z $\sim$ 1
based on a comparison of the total amount of star-formation between
z $=$ 0 and z $=$ 1.3 (cf. Madau, this volume) and the upper
limits on the brightness of the cosmic background ionizing-radiation field
in the present universe. 

Thus, the role of starbursts (and high-mass stars in general) in
photoionizing the IGM is still unclear. The discovery of star-
forming galaxies at z $>$ 3 in principle allows the spectral region
below the Lyman-break to be probed directly with the new generation
of 8 and 10-meter-class telescopes, and so the relative importance
of high-mass stars and QSOs in photoionizing the IGM can be
determined for this early epoch.

\subsection*{Galactic Superwinds and the `Pollution' of the IGM}

Over the last few years, observations have provided convincing
evidence of the existence (and even the ubiquity) of `superwinds' -
 galactic-scale outflows of gas driven by the collective effect of
multiple supernovae and stellar winds in a starburst (cf. Heckman,
Lehnert, \& Armus 1993; Lehnert \& Heckman 1996a; Veilleux, Cecil, \& Hawthorn 
1996). X-ray data have proved particularly crucial since they
are the only direct probe of the hot gas that contains most of the
mass and energy in the flow.

Soft X-ray emission (hot gas) is a generic feature of the halos of
the nearest starburst galaxies (Dahlem, Weaver, \& Heckman 1997).
The estimated thermal energy content of this
gas represents a significant fraction of the time-integrated
mechanical energy supplied by the starburst, while the soft X-ray
luminosity is only a few \% of the supernova heating rate. These
results mean that little of the mechanical energy supplied by
supernovae and stellar winds in starbursts is 
radiated away. Thus, in principle superwinds may efficiently
transport much of the mechanical energy supplied by high-mass stars
into the IGM, making such stars an important heating source for the
IGM (see Donahue's contribution to this volume).

The temperature of the hot outflowing gas in starbursts is
considerably cooler (few to ten million degrees) than would be
expected for pure thermalized supernova+stellar wind ejecta
(10$^{8}$ K). Given these temperatures, will this gas escape the
galaxy gravitational potential, or will the gas remain bound and
perhaps cool and return as a galactic fountain?  Following Wang
(1995), the `escape temperature' for hot gas in a galaxy potential
with an escape velocity v$_{esc}$ is given by
T$_{esc}$ $\sim$ 4 $\times$ 10$^{6}$ (v$_{esc}$/600 km s$^{-
1}$)$^{2}$ K (where I have roughly chosen parameters appropriate
to a typical spiral galaxy like our own).

The gas temperatures in superwinds are observed to be independent
of the galaxy rotation speed, while T$_{esc}$ should scale as the
square of the rotation speed ({\it modulo} the extent of the dark
matter halo). Thus, it appears that the X-ray emitting gas can
easily escape from dwarf galaxies undergoing starbursts (Della-Ceca
et al 1996) but possibly not from the halos of the most massive
starburst galaxies (Dahlem, Weaver, \& Heckman 1997; Wang 1995).
As Larson \& Dinerstein (1975) and many others have proposed, this
selective loss of metal-enriched gas from the shallowest galaxy
potential wells may be the physical mechanism that underlies the
strong correlation between the metal abundance of the stellar
population and the escape velocity in elliptical galaxies (e.g.
Franx \& Illingworth 1990). In this sense, superwinds should have
a particularly devastating impact on dwarf galaxies (e.g. Dekel \&
Silk 1986; Marlowe et al 1995).

If indeed superwinds carry substantial amounts of metal-enriched
gas out of starbursts, we should see the cumulative effect of these
flows in the form of a metal-enriched IGM and/or metal-enriched
gaseous halos around galaxies. By now it is clear that typical 
MgII absorption-line systems at z $<$ 1 seen in the
spectra of distant QSOs arise in the metal-enriched halos of intervening
galaxies (cf. Churchill, Steidel, \& Vogt 1996). These galaxies
appear to be normal systems (not starbursts), so it is unlikely that
`living, breathing' superwinds are implicated. Nevertheless, it is
plausible that the metals in these galaxy halos may trace
`fossil' superwinds. That is, galactic halos may be polluted primarily by the
episodic eruptions associated with powerful starbursts (Heckman 1997).

Can we say anything about the gas {\it outside} galaxies? The existence
of an IGM in clusters of galaxies whose metal content exceeds that
of all the stars in all the cluster's galaxies is one of the most
remarkable phenomena in extragalactic astronomy (see Donahue,
this volume). Recent ASCA X-ray spectra of this gas
show super-solar abundance ratios for the $\alpha$-process
elements like O, Ne, and Si relative to Fe (Mushotzky et al 1996).
This implicates `core-collapse' supernovae (the end product of
high-mass stars) and - by inference - superwinds
as the source of the metals in the cluster IGM (Loewenstein \&
Mushotzky 1996).

If most of the metals in clusters of galaxies are floating around
outside galaxies, could this be true more globally in the universe?
We don't know the answer in the present-day universe, but the
situation at high-redshift is very intruiging. 
Burles \& Tytler (1996) use the detection of OVI absorption-lines
in the spectra of background QSOs to estimate that the minimum metallicity
of the entire gas-phase baryonic component of the universe at z $\sim$ 1
is $>$ 0.02 h solar for $\Omega_{B,gas} =$ 0.01
h$^{-2}$. This is a lower limit, because it assumes all the Oxygen is the
form of OVI and that the detected OVI absorption-lines
are optically-thin. At z $>$ 2,
the presence of metals in the `Ly$\alpha$ forest'
(the `cloudy' component of the early IGM) is
certainly suggestive of the dispersal of chemically-enriched
material by early superwinds (cf. Cowie et al 1995; Tytler et al
1995; Madau \& Shull 1996). The Ly$\alpha$ forest material appears to
have a metallicity of about 10$^{-2}$ solar, but a
high ratio of Si/C that is suggestive of core-collapse supernovae
as the source of metals (Cowie et al 1995; Giroux \& Shull 1997).
Given the above indications of a metal-enriched IGM at high-z, do we see
any {\it direct} evidence for the outflow of metal-enriched gas from
starburst galaxies in the early universe? We will return briefly to this in
section 4.2 below.

The relatively large inferred masses of the X-ray emitting gas in
local superwinds imply that we are primarily observing emission from
ambient gas that has been `mass-loaded' in some way into the
superwind (cf. Suchkov et al 1996; Hartquist et al 1997;
Della Ceca et al 1996). That is,
each supernova explosion on-average must heat and eject a few
hundred M$_{\odot}$ worth of gas up to of-order 10$^{7}$ K. The
large amount of mass-loading implies that the ratio of gas that is
blown out of the starburst compared to the mass that is turned into
stars ranges from of-order unity (if a normal complement of low-
mass stars is formed in the starburst) to of-order ten (if only
high-mass stars are formed). To the extent that local starbursts
are analogs to forming galaxies, this suggests that galaxy
formation may have been an inefficient process with only a minority
of the initial complement of baryons being retained and converted
into stars and the mAJority expelled into galactic halos or the
IGM.
\newpage

\section*{Starbursts in the Ultraviolet}

\subsection*{Local Starbursts as a `Training Set'}

Observations in the vacuum-UV spectral regime ($\lambda$ $\sim$ 912
to 3000 \AA) are crucial for both understanding local starbursts,
and for relating them to galaxies at high-redshift. Not only is
this the energetically-dominant spectral region for the hot stars
that power the starburst, this is the spectral regime where we can
most clearly observe the direct spectroscopic signatures of these
hot stars. Moreover, the vacuum-UV contains a wealth of spectral
features including the resonance transitions of most cosmically-
abundant ionic species. These give UV spectroscopy a unique
capability for diagnosing the (hot) stellar population and the
physical and dynamical state of gas in starbursts. 

Since ground-based optical observations of galaxies at high-
redshifts sample the vacuum-UV portion of their rest-frame
spectrum, we can not understand how galaxies evolved without
documenting the vacuum-UV properties of galaxies in the present
epoch. In particular, a thorough understanding of how to exploit
the diagnostic power of the rest-frame UV spectral properties of
local starbursts will give astronomers powerful tools with which
to study star-formation and galaxy-evolution in the early universe.

The vacuum-UV spectra of starbursts are characterized by strong
absorption features, as seen in Figure 1.
These absorption features can have three
different origins: stellar winds, stellar photospheres, and
interstellar gas. Detailed analyses of Hubble Space Telescope (HST)
and Hopkins Ultraviolet Telescope (HUT) spectra of starbursts show
that the resonance lines due to species with low-ionization
potentials (OI, CII, SiII, FeII, AlII, etc.) are primarily
interstellar in origin. In contrast, the resonance lines due to
high-ionization species (NV, SiIV, CIV) can contain significant
contributions from both stellar winds and interstellar gas, with
the relative importance of each varying from starburst to starburst
(cf. Conti et al 1996; Leitherer et al 1996; Heckman \& Leitherer
1997; Gonzalez-Delgado et al 1997). The most unambiguous detection
of stellar photospheric lines is provided by excited transitions,
but these lines are usually rather weak (cf. Heckman \& Leitherer 1997).

\vskip10truept

\begin{figure}
\centerline{\epsfig{figure=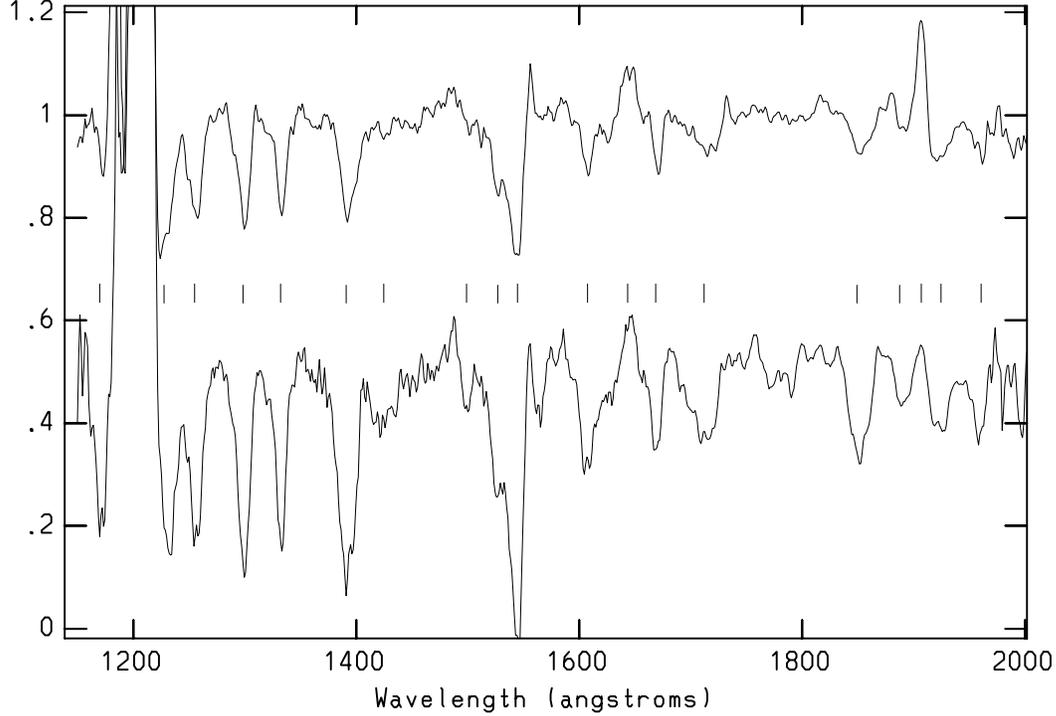}}
\vspace{15pt}
\caption{IUE spectra of local starbursts with low-metallicity
(top) and high-metallicity (bottom). Each spectrum is a weighted
average of the spectra of about 20 starbursts. The mean metallicities
are 0.2 solar (top) and 1.3 solar (bottom). A number of features
are indicated by tick marks and have the following identifications
(from left-to-right): CIII$\lambda$1175 (P), NV$\lambda$1240 (W),
SiII$\lambda$1260 (I), OI$\lambda$1302 plus SiII$\lambda$1304 (I),
CII$\lambda$1335 (I), SiIV$\lambda$1400 (W;I), SiIII$\lambda$1417 plus
CIII$\lambda$1427 (P), SV$\lambda$1502 (P), SiII$\lambda$1526 (I),
CIV$\lambda$1550 (W;I),
FeII$\lambda$1608(I), HeII$\lambda$1640 emission (W), AlII$\lambda$1671 (I),
NIV$\lambda$1720 (W), AlIII$\lambda$1859 (I;W), SiIII$\lambda$1892 (P),
CIII]$\lambda$1909 (nebular emission-line), FeIII$\lambda$1925 (P),
FeIII$\lambda$1960 (P). Here, I, P, and W denote
lines that are primarily of interstellar, stellar photospheric, or stellar
wind origin. The strong emission feature near 1200 \AA\ is geocoronal
Ly$\alpha$.}
\end{figure}

We (Heckman et al 1997 - hereafter H97) have just completed an
analysis of the vacuum-UV spectroscopic properties of a large
sample of starburst galaxies in the local universe using the data
archives of the International Ultraviolet Explorer (IUE) satellite.
Taken together with the results of previous studies of starbursts in the
vacuum-UV, the principal lessons we have learned are as follows:\\[1mm]

{\bf First, dust has a profound effect on the emergent UV
spectrum.}

As noted above, the luminous output of a starburst should have its
peak in the vacuum-ultraviolet range ($\lambda$ $\sim$ 912 to 3000
\AA). On the other hand, starbursts are rich in interstellar gas
and dust, and the effective absorption cross-section of these dust
grains is a strong inverse function of wavelength. This means that
the effect of dust on the vacuum-UV properties of starbursts is
profound.

Previous papers have established that various independent
indicators of dust extinction in starbursts observed with IUE
correlate strongly with one another. Calzetti et al (1996) show
that the spectral slope in the vacuum-UV continuum (as
parameterized by $\beta$, where F$_{\lambda} \alpha\
\lambda^{\beta}$) correlates strongly with the nebular extinction
measured in the optical using the Balmer decrement. Meurer et al
(1995;1997) show that $\beta$ correlates well with the ratio of
far-IR to vacuum-UV flux: the greater the fraction of the UV that
is absorbed by dust and re-radiated in the far-IR, the redder the
vacuum-UV continuum. The interpretation of these correlations with
$\beta$ in terms of the effects of dust are particularly plausible
because the {\it intrinsic} value for $\beta$ in a starburst is a
robust quantity. Figures 31 and 32 in Leitherer \& Heckman (1995)
show that $\beta$ should have a value between about -2.0 and -2.6
for the range of ages and initial mass functions appropriate for
starbursts (cf. Leitherer 1997).

Our detailed understanding of the above results is incomplete,
since they must involve both the geometrical distribution of the
dust, stars, and gas in the starburst and the vacuum-UV extinction
law for the dust. However, the available data strongly suggest that
(quite surprisingly) much of the dust responsible for the vacuum-
UV extinction is apparently distributed around the starburst in the
form of a moderately inhomogeneous foreground screen or `sheath'
surrounding the starburst (Gordon, Calzetti, \& Witt 1997).

Interestingly, H97 find that the amount of vacuum-UV extinction in
starbursts correlates strongly with the bolometric luminosity of
the starburst: only starbursts with L$_{bol}$ $<$ few $\times
10^{9}$ L$_{\odot}$ have colors expected for an unreddened
starburst and have vacuum-UV luminosities that rival their far-IR
luminosities. Starbursts that lie at or above the `knee' in the
local starburst luminosity function (L$_{bol}$ $>$ few $\times
10^{10}$ L$_{\odot}$ - cf. Soifer et al 1987) have red UV continua
($\beta \sim$ -1 to +0.4) and are dominated by far-IR emission
(L$_{IR}$ $\sim$ 10 to 100 L$_{UV}$). We also find that the amount
of vacuum-UV extinction in starbursts correlates well with the
absolute blue magnitude and the rotation speed of the galaxy
`hosting' the starburst: starbursts in more massive galaxies are
more dust-shrouded.\\[1mm]

{\bf Second, the metallicity of the starburst also strongly affects
the UV spectrum.}

Apart from the effects of dust, a starburst's metallicity is the
single most important parameter in determining its vacuum-UV
properties. In fact, metallicity and the effects of dust are well-
correlated, as shown in Figure 2.
\vskip10truept

\begin{figure}
\centerline{\epsfig{figure=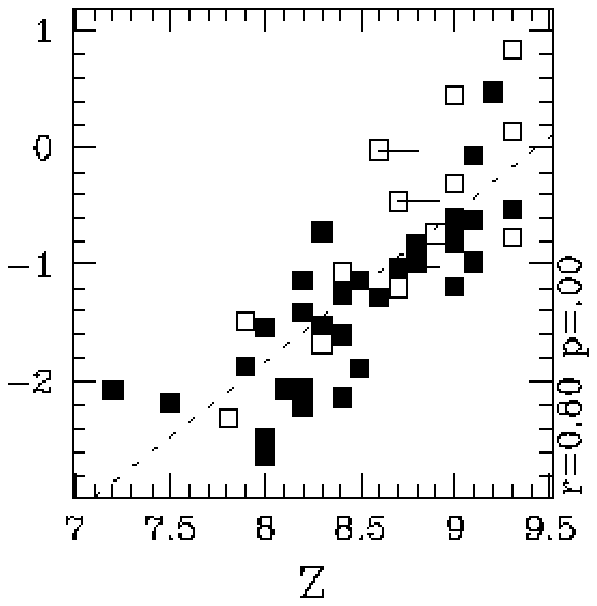}
\epsfig{figure=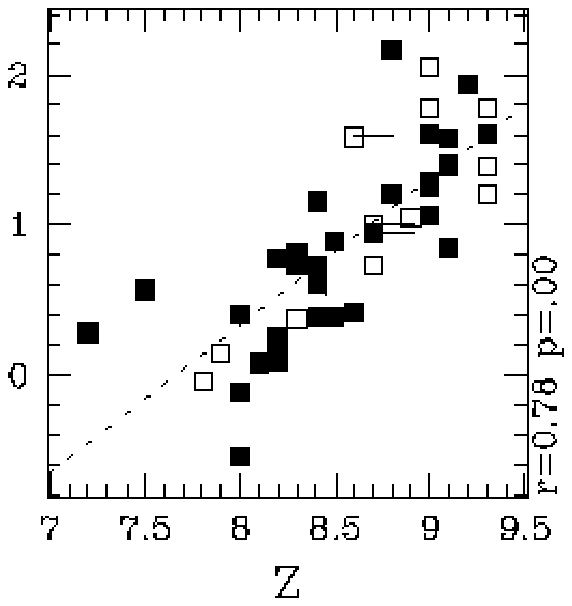}}
\vspace{15pt}
\caption{Plots of the log of the starburst metallicity (on a scale
where the solar value is 8.9) {\it versus} two dust-indicators:
the spectral slope of the vacuum-UV continuum (as
parameterized by $\beta$, where F$_{\lambda} \alpha\
\lambda^{\beta}$) and the log of the ratio of far-IR to vacuum-UV flux.
Heavily-reddened and -extincted metal-rich starbursts lie to the upper
right of each plot.}
\end{figure}

At low metallicity ($<$10\% solar) a significant fraction of the
intrinsic vacuum-UV actually escapes the starburst
(L$_{IR}$/L$_{UV}$ $\sim$ unity), and the vacuum-UV colors are
consistent with the intrinsic (unreddened) colors expected for a
starburst population ($\beta \sim$ -2). In contrast, at high
metallicities ($>$ solar) 90\% to 99\% of the energy emerges
in the far-IR (L$_{IR}$/L$_{UV}$ = 10 to 100) and the vacuum-UV
colors are very red ($\beta \sim$ 0). Storchi-Bergmann, Calzetti,
\& Kinney (1994) had previously noted the correlation between
metallicity and UV color. 

These correlations have a straightforward interpretation: the
vacuum-UV radiation escaping from starbursts suffers an increasing
amount of reddening and extinction as the dust-to-gas ratio in the
starburst ISM increases with metallicity. This will be true
provided that neither the gas column density towards the starburst,
nor the fraction of interstellar metals locked into dust grains are
strong inverse functions of metallicity.

The properties of the vacuum-UV absorption-lines are also strongly
dependent on metallicity. Figure 1 shows that both the
high-ionization
(e.g. CIV$\lambda$1550 and SiIV$\lambda$1400) and low-ionization
(e.g. CII$\lambda$1335, OI$\lambda$1302, and
SiII$\lambda$$\lambda$1260,1304) resonance absorption-lines arei
significantly stronger in starbursts with high metallicity.

The metallicity-dependence of the high-ionization lines (noted
previously by Storchi-Bergmann, Calzetti, \& Kinney 1994) is not
surprising, given the likely strong contribution to these lines
from stellar winds. Theoretically, we expect that since stellar
winds are radiatively driven, the strengths of the vacuum-UV
stellar wind lines will be metallicity-dependent. This is confirmed
by available HST and HUT spectra of LMC and especially SMC stars
(Walborn et al 1995; Puls et al 1996). If post-main-sequence stars
(supergiants and Wolf-Rayet stars) contribute significantly to the
integrated light, the stellar wind properties enter in a second,
indirect way. That is, the integrated UV spectrum heavily depends
on the evolutionary history of the OB population, which in turn is
critically dependent on the stellar mass-loss rates, which in turn
are a function of the metal abundance (see Maeder \& Conti 1994).

Figure 1 also shows a metallicity-dependence for the strengths of the UV
absorption-lines that are of stellar-photospheric rather than
interstellar origin (we know they are photospheric lines because they
correspond to transitions out of
highly excited states). Such lines are generally rather weak in starburst
spectra and/or blended with strong interstellar features. They
include CIII$\lambda$1175,
SiIII$\lambda$1417, CIII$\lambda$$\lambda$1426,1428,
SV$\lambda$1502, SiIII$\lambda$1892, and
FeIII$\lambda\lambda$1925,1960.

The weak but statistically-significant correlation between
metallicity and the strength of the low-ionization resonance lines
(which are primarily formed in the interstellar medium of the
starburst) is also unsurprising. Analyses of HST spectra (cf. Pettini \& 
Lipman 1995;
Heckman \& Leitherer 1997; Sahu \& Blades 1997; Gonzalez-Delgado
et al 1997) show that the strong interstellar lines are saturated
(highly optically-thick). In this case, the equivalent width of the
absorption-line (W) is only weakly dependent on the ionic column
density (N$_{ion}$): W $\alpha$\ b[ln(N$_{ion}$/b)]$^{0.5}$, where
b is the normal Doppler line-broadening parameter. Over the range
that H97 sample well, the starburst metallicity increases by a
factor of almost 40 (from 0.08 to 3 solar), while the equivalent
widths of the strong interstellar lines only increase by an average
factor of about two. This is consistent with the strong
interstellar lines being quite optically-thick.\\[1mm]

{\bf Third, the properties of the strong interstellar absorption-
lines reflect the hydrodynamical consequences of the
starburst, and do not straightforwardly probe the
gravitational potential of the galaxy.}

As noted above, analyses of HST and HUT UV spectra of starbursts
imply that the interstellar absorption-lines lines are optically-
thick. Their strength is therefore determined to first-order by the
velocity dispersion in the starburst (see above). Thus, these lines
offer a unique probe of the kinematics of the gas in starbursts.
The enormous strengths of the starburst interstellar lines
(equivalent widths of 3 to 6 \AA\ in metal-rich starbursts) require
very large velocity dispersions in the absorbing gas (few hundred
km s$^{-1}$). Are these gas motions primarily due to gravity or to
the hydrodynamical `stirring' produced by supernovae and stellar
winds?

Both processes probably contribute to the observed line-broadening.
H97 find only a very weak (but still
statistically significant) correlation between the strengths
(widths) of the interstellar absorption-lines and the rotation-
speed of the host galaxy. The weakness of the correlation suggests
that gravity alone is not the whole story. Moreover, for any
astrophysically-plausible ionic column density, the interstellar
lines are typically at least twice as broad as can be explained by
optically-thick gas orbiting in the galaxy
potential well.

The most direct evidence for a non-gravitational origin of the gas
motions comes from analyses of HST and HUT
spectra, which show that the interstellar lines are often
blueshifted by one-to-several-hundred km s$^{-1}$ with respect to
the systemic velocity of the galaxy (Heckman \& Leitherer 1997;
Gonzalez-Delgado et al 1997; Sahu \& Blades 1997; Lequeux et al 1995).
This demonstrates directly that the absorbing gas is flowing
outward from the starburst, probably `feeding' the superwinds
described in section 3.2 above).

\subsection*{Implications at High-Redshift}

The results summarized in section 4.1 have a variety of interesting
implications for the interpretation of the rest-frame-UV properties
of galaxies at high-redshift.

Powerful starbursts in the present universe emit almost all their
light in the far-infrared, not in the ultraviolet. Thus, an
ultraviolet census of the local universe would significantly
underestimate the true star-formation-rate and would systematically
under-represent the most powerful, most metal-rich starbursts
occuring in the most massive galaxies.
This may also be true at
high-redshift, where the current estimates of star-formation rely
almost exclusively on data pertaining to the rest-frame vacuum-UV
(see Madau, this volume). For example, current samples might under-
represent young/forming massive elliptical galaxies.

Using the strong correlation between the vacuum-UV color of local
starbursts ($\beta$) and the ratio of far-IR to vacuum-UV light
emitted by local starbursts, Meurer et al (1997) estimate that an
average vacuum-UV-selected galaxy at high-redshift (e.g. Steidel
et al 1996; Lowenthal et al 1997) suffers 2 to 3 magnitudes of
extinction. The `correct' prescription for de-extincting the high-
z galaxies in order to correctly obtain the bolometric luminosity
and star-formation rate is a matter of on-going debate (see the
contributions by Pettini and Madau to this volume). It is worth
emphasizing that the purely empirical method of Meurer et al
bypasses the substantial uncertainties about the dust extinction
law, and the initial mass function and age of the stellar
population in the high-z galaxies (it assumes only that the high-z
galaxies behave like local starbursts).

As shown recently by Burigana et al (1997), the existing limits on
the far-IR/sub-mm cosmic background are consistent with the global
star-formation rates inferred by Meurer et al at z$>$2 due to dusty
starbursts unless the dust in these galaxies is quite cool
(T$_{dust}$ $<$ 20 K) compared to the dust in local starbursts
(T$_{dust}$ $\sim$ 30
to 60 K). This seems very unlikely, since the bolometric surface-brightnesses
of the high-redshift galaxies are similar to local starbursts (Meurer et al
1997), implying that the energy density of the radiation field that heats
the grains is similar in the two types of objects (cf. Figure 5 and the 
relevant discussion in Lehnert \& Heckman (1996b).

In any case, it seems fair to conclude that
the history of star-formation in the universe at early times (z $>$
1) will remain uncertain until the effects of dust extinction are
better understood.

The strong correlation shown in Figure 2
between vacuum-UV color ($\beta$) and
metallicity in local starbursts - if applied naively to high-z
galaxies - would suggest a broad range in metallicity from
substantially subsolar to solar or higher and a median value of
0.3 to 0.5 solar. This is signifcantly higher than the mean
metallicity in the damped Ly$\alpha$ systems (the mAJor repository
of HI gas at these redshifts), but this may be due to selection
effects: the UV-selected galaxies are the most actively star-
forming regions of galaxies, while the damped Ly$\alpha$ systems
tend to sample the outer, less-chemically-enriched parts of
galaxies or perhaps proto-galactic fragments (e.g. Pettini et al 1997).

It would be interesting to use the correlations between absorption-
line strengths and metallicity in local starbursts to `guesstimate'
the metallicity of the high-z galaxies (see Pettini, this volume).
One prediction based on the local starbursts (H97) is that the
high-z galaxies should show a strong correlation between the
strength of the UV absorption-lines (stellar and interstellar) and
$\beta$ (the more metal-rich local starbursts are both redder and
stronger-lined).

As noted above, Meurer et al (1997) argue that the UV-selected
galaxies at high-redshift suffer substantial amounts of extinction.
If their proposed extinction-corrections are applied, the high-z
galaxies have very large bolometric luminosities ($\sim$ 10$^{11}$
to 10$^{13}$ L$_{\odot}$ for H$_{0}$ = 75 km s$^{-1}$ Mpc$^{-1}$
and q$_{0}$ = 0.1). Interestingly, the bolometric surface-
brightnesses of the extinction-corrected high-z galaxies are very
similar to the values seen in local starbursts: $\sim$ 10$^{10}$
to 10$^{11}$ L$_{\odot}$ kpc$^{-2}$. The high-redshift galaxies
appear to be `scaled-up' (larger and more luminous) versions of the
local starbursts. The physics behind this `characteristic' surface-
brightness is unclear (cf. Meurer et al 1997; Lehnert \& Heckman
1996b). However, it is intruiging that the implied average surface-
mass-density of the stars within the half-light radius ($\sim$
10$^{2}$ to 10$^{3}$ M$_{\odot}$ pc$^{-2}$) is quite similar to the
values in present-day elliptical galaxies. Are we witnessing the
formation of elliptical and bulges?

Finally, based on local starbursts - it seems likely that the gas
kinematics that are measured in the high-z galaxies using the
interstellar absorption-lines are telling us a great deal about the
hydrodynamical consequences of high-mass star-formation on the
interstellar medium, but rather little (at least directly) about
the gravitational potential or mass of the galaxy. Even the widths
of the nebular emission-lines in local starbursts are not always
reliable tracers of the galaxy potential well (cf. Lehnert \&
Heckman 1996b). This means that it will be tricky to determine
masses for the high-z galaxies without measuring real rotation-
curves via spatially-resolved spectroscopy.

On the brighter side, if the kinematics of the interstellar
absorption-lines can be generically shown to arise in outflowing
metal-enriched gas, we can then directly study high-redshift star-
forming galaxies caught in the act of `polluting' the intra-cluster
medium and inter-galactic medium with metals in the early universe
(see section 3.2 above).

In fact, there is now rather direct observational evidence that
this is the case. I will need to briefly digress to explain this
evidence. As emphasized above, the interstellar absorption-lines
are significantly blue-shifted with respect to the systemic
velocity of the galaxy (v$_{sys}$) in many local starbursts. In the
high-redshift galaxies there is rarely a good estimator of
v$_{sys}$ (although the weak stellar photospheric lines listed in
section 4.1 above are a promising possibility in spectra with
adequate signal-to-noise). It is also the case in local starbursts
that the true galaxy systemic velocity lies between the velocity
of the UV interstellar absorption-lines and the Ly$\alpha$ {\it
emission} line (Lequeux et al 1995; Gonzalez-Delgado et al 1997).
This is due to outflowing gas that both produces the blue-shifted
absorption-lines and absorbs-away the blue side of the Ly$\alpha$
emission-line. Thus, a purely-UV signature of outflowing gas is a
blueshift of the interstellar absorption-lines with respect to the
Ly$\alpha$ emission-line (even though neither is at v$_{sys}$).

\begin{figure}
\centerline{\epsfig{figure=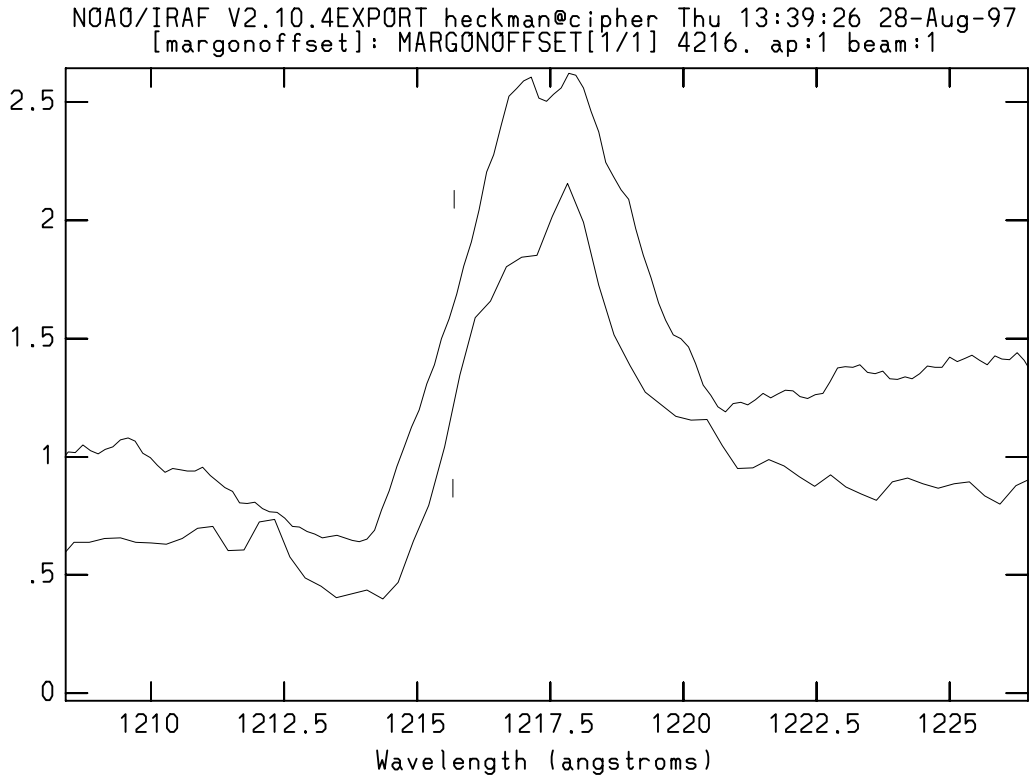}}
\vspace{10pt}
\caption{Overplot of an HST spectrum of the local starburst IRAS 0833+6517
(top - with an arbitrary offset along the y-axis)
and the composite spectrum of 12 high-redshift galaxies. The rest-frame
wavelength of Ly$\alpha$ at v$_{sys}$ in IRAS 0833+6517 is indicated by
the tick marks. Note the strong similarity of the two spectra with
redshifted Ly$\alpha$ emission and blueshifted
absorption, both due to outflowing gas.
See text for details}.
\end{figure}

Recently, Franx et al (1997) have seen just this effect in a spectrum
of the highest-redshift object in the universe: a 
gravitationally-lensed galaxy at z = 4.92. They find that the
Ly$\alpha$ emission line is redshifted by about 400 km s$^{-1}$
relative to the SiII$\lambda$1260 interstellar absorption-line
across the entire face of the galaxy. More generally, Lowenthal et
al (1997) have constructed a composite UV spectrum of 12 high-z
galaxies. In Figure 3 I overplot this spectrum on an HST spectrum
of the local starburst IRAS 0833+6517 (Gonzalez-Delgado et al 1997).
The spectra have been aligned to force a coincidence in wavelength
between the strong interstellar absorption-lines. The similarity of the
two spectra is striking: a strong redshifted Ly$\alpha$ emission-
line and a weak blue-shifted Ly$\alpha$ {\it absorption} line. This
composite spectrum strongly suggests that the outflow of metal-
enriched gas at velocities of a few hundred km s$^{-1}$ is a
generic feature of the high-z galaxies. {\it If} the outflowing gas
escapes into the IGM, such flows could bring an IGM with
$\Omega_{IGM}$ $\sim$ 0.01 h$^{-2}$ up to a mean metallicity of
$>$ 10$^{-2}$ solar by a redshift of 2.5 (cf. Madau \& Shull 1996).

\section*{Summary}

Starbursts are defined as brief episodes ($<$ 10$^{8}$ years) of
intense star-formation that occur in the central-most 0.1 to 1 kpc-
scale regions of galaxies and dominate the integrated emission from
the galaxy. They are a significant component of the present-day
universe: they provide roughly 10\% of the bolometric emissivity
of the local universe and are the sites of $\sim$25\% of the high-
mass star-formation. Thus, they deserve to be understood in their
own right.

They also offer unique `laboratories' for testing our ideas about
star formation, the evolution of high-mass stars, and the physics
of the interstellar medium. They serve as local analogs of the
processes that were important in the origin and early evolution of
galaxies and in the heating and chemical enrichment of the inter-
galactic medium (IGM).

In this contribution I have reviewed starbursts from this broad
cosmogonical perspective, stressing several key lessons we have
learned from starbursts:\\[1mm]

{\bf i) Violent, transient events play a significant role in the
origin and evolution of galaxies.} Rather than simply evolving in
a steady `clockwork' fashion, galaxies also evolve discontinuously
through powerful bursts of star-formation, triggered by galaxy
interactions.\\[1mm]

{\bf ii) Galaxies do not evolve as `Island Universes' or `Closed
Boxes.'} Powerful starbursts are triggered by galaxy interactions
and mergers that can create an elliptical galaxy out of two
spirals. The starbursts produce outflows of hot metal-enriched gas
(`superwinds') that pollute the inter-galactic medium. There is now
direct observational evidence for a metal-enriched IGM and for the
outflows at low- and high-redshift that are responsible for this
enrichment. Study of superwinds in the local universe suggest that
galaxy formation may have been an inefficient process with the
mAJority of gas being ejected.\\[1mm]

{\bf iii) Dust dramatically affects of view of high-mass star-
formation in starbursts and (probably) in high-redshift galaxies.} 
Powerful starbursts in the present universe emit almost all their
light in the far-infrared. An ultraviolet census of the local
universe would significantly underestimate the true star-formation-
rate and could systematically under-represent the most powerful,
most metal-rich starbursts occuring in the most massive galaxies.
Applying the empirical relations followed by local starbursts to
the UV-selected galaxies at high-redshift implies that the latter
typically suffer 2 to 3 magnitudes of extinction, and that the
star-formation rate in the universe need not be smaller at z$\sim$
3 than at z$\sim$1. The high-z galaxies would then have estimated bolometric
surface-brightnesses and sizes consistent with a population of
forming elliptical galaxies.\\[1mm]

{\bf iv) Space-based observations are crucial in understanding starbursts.}
X-ray data (ROSAT,
ASCA, AXAF, XMM) are vital for studying the superwind phenomenon, since
they directly probe the hot gas that comprises most of the mass and
energy in the flow. Ultraviolet data (IUE, HST, HUT, FUSE) offer
detailed information about both the hot stars that power the
starburst and the dynamics of the interstellar medium. Visible
observations with the high-angular resolution of HST have revealed
the importance of galaxy interactions and mergers at low and medium
redshift and given us a tantalizing glimpse of the star-forming
history of the early universe. Finally, the infrared (IRAS, ISO,
COBE, WIRE, SIRTF, SOFIA, NGST) represents the primary channel for
energy loss for the dusty, metal-rich starbursts. Such data not
only document the presence of dusty starbursts in the local
universe (z $<<$ 1), but also constrain their global importance at
high-redshift.

\section*{Acknowledgments}

I would like to thank my collaborators over the past few years
(especially D. Calzetti, M. Dahlem, R. Della Ceca, R. Gonzalez-
Delgado, M. Lehnert, C. Leitherer, G. Meurer, C. Robert, A.
Suchkov, and K. Weaver) for hard work and inspiration. Special
thanks to J. Lowenthal for providing a composite spectrum of high-
z galaxies and to M. Carollo, M. Dickinson, P. Madau, C. Martin,
C. Mihos, C. Norman, M.
Pettini, C. Steidel, and M. Voit for stimulating discussions and
constructive suggestions. This work has been partially supported
by NASA grant NAGW-3138.

\end{document}